\documentclass[fleqn,usenatbib]{mnras}
\usepackage{newtxtext,newtxmath}

\usepackage[T1]{fontenc}
\usepackage{ae,aecompl}
\usepackage{chngcntr}

\defcitealias{2018MNRAS.477.4380S}{S18}
\defcitealias{2018Natur.560..613T}{T18}
\defcitealias{1998ApJ...498..541K}{KS}
\defcitealias{2012ApJ...745...69K}{KDM}
\defcitealias{2015ApJ...806L..36S}{SFK}
\usepackage{etoolbox}
\makeatletter
\patchcmd\@combinedblfloats{\box\@outputbox}{\unvbox\@outputbox}{}{%
   \errmessage{\noexpand\@combinedblfloats could not be patched}%
}%
 \makeatother

\usepackage{graphicx}	
\usepackage{amsmath}	
\usepackage{amssymb}	

\newcommand{\plasmabeta}{\beta_{\mathrm{mag}}}

\newcommand{\mach}{\mathcal{M}}

\newcommand{\sfrunits}{\msol\,\mathrm{yr}^{-1}\,\mathrm{kpc}^{-2}}

\newcommand{\msol}{\mbox{$\mathrm{M}_{\sun}$}}
\newcommand{\lsol}{\mbox{$\mathrm{L}_{\sun}$}}

\newcommand{\tff}{t_\mathrm{ff}}

\newcommand{\pc}{\mathrm{pc}}
\newcommand{\kpc}{\mathrm{kpc}}

\newcommand{\yr}{\mathrm{yr}}

\newcommand{\siggas}{\Sigma_\mathrm{gas}}

\newcommand{\sigmasfr}{\Sigma_{\mathrm{SFR}}}

\title[Testing Star Formation Relations on AzTEC-1]{Testing Star Formation Laws on Spatially Resolved Regions in a $z \approx 4.3$ Starburst Galaxy}

\author[P. Sharda et al.]{
P. Sharda$^{1,2}$\thanks{E-mail: piyush.sharda@anu.edu.au (PS)},
E. da Cunha$^{1}$\thanks{E-mail: elisabete.dacunha@anu.edu.au (EdC)}, 
C. Federrath$^{1}$,
E. Wisnioski$^{1,2}$,
E. M. Di Teodoro$^{1,2}$,
\newauthor
K. Tadaki$^{3}$,
M. S. Yun$^{4}$,
I. Aretxaga$^{5}$, and
R. Kawabe$^{3,6,7}$\\
$^{1}$Research School of Astronomy and Astrophysics, Australian National University, Canberra, ACT 2611, Australia\\
$^{2}$ARC Centre of Excellence for All Sky Astrophysics in 3 Dimensions (ASTRO 3D), Australia\\
$^{3}$National Astronomical Observatory of Japan, 2-21-1 Osawa, Mitaka, Tokyo 181-8588, Japan\\
$^{4}$Department of Astronomy, University of Massachusetts, Amherst, MA 01003, USA\\
$^{5}$Instituto Nacional de Astrof\'isica, Optica y Electr\'onica (INAOE), Aptdo. Postal 51 y 216, 72000 Puebla, Mexico\\
$^{6}$SOKENDAI (The Graduate University for Advanced Studies), 2-21-1 Osawa, Mitaka, Tokyo 181-0015, Japan\\
$^{7}$Department of Astronomy, School of Science, University of Tokyo, Bunkyo, Tokyo 113-0033, Japan\\
}
\date{Accepted XXX. Received YYY; in original form ZZZ}

\pubyear{2019}

\begin{document}
\label{firstpage}
\pagerange{\pageref{firstpage}--\pageref{lastpage}}
\maketitle

\begin{abstract}
We probe the star formation properties of the gas in AzTEC-1 in the COSMOS field, one of the best resolved and brightest starburst galaxies at $z \approx 4.3$, forming stars at a rate $>\,1000\,\msol\,\mathrm{yr^{-1}}$. Using recent ALMA observations, we study star formation in the galaxy nucleus and an off-center star-forming clump and measure a median star formation rate (SFR) surface density of $\Sigma^{\mathrm{nucleus}}_{\mathrm{SFR}} = 270\pm54$ and $\Sigma^{\mathrm{sfclump}}_{\mathrm{SFR}} = 170\pm38\,\msol\,\mathrm{yr}^{-1}\,\mathrm{kpc}^{-2}$, respectively. Following the analysis by Sharda et al. (2018), we estimate the molecular gas mass, freefall time and turbulent Mach number in these regions to predict $\Sigma_{\mathrm{SFR}}$ from three star formation relations in the literature. The Kennicutt-Schmidt (Kennicutt 1998, KS) relation, which is based on the gas surface density, underestimates the $\Sigma_{\mathrm{SFR}}$ in these regions by a factor 2-3. The $\Sigma_{\mathrm{SFR}}$ we calculate from the single-freefall model of Krumholz et al. 2012 (KDM) is consistent with the measured $\Sigma_{\mathrm{SFR}}$ in the nucleus and the star-forming clump within the uncertainties. The turbulence-regulated star formation relation by Salim et al. 2015 (SFK) agrees slightly better with the observations than the KDM relation. Our analysis reveals that an interplay between turbulence and gravity can help sustain high SFRs in high-redshift starbursts. It can also be extended to other high- and low-redshift galaxies thanks to the high angular resolution and sensitivity of ALMA observations.
\end{abstract}




\section{Introduction}
\label{s:intro}
Understanding the formation and evolution of stars in the Universe remains one of the most pertinent questions in astrophysics. Deep surveys have established that the epoch of maximum star formation corresponds to redshifts $1 < z < 3$ \citep{2014ARA&A..52..415M}. As more and more starburst galaxies are found at $z\,\ga\,4$ with star formation rates (SFRs) exceeding $1000\,\msol\,\mathrm{yr^{-1}}$ (e.g., \citealt{2010MNRAS.407L.103C,2010ApJ...709..210K,2016A&A...586L...7B,2018ApJ...861...43P}), it may imply that there is a higher fraction of them than previously estimated \citep{2006MNRAS.370..645B,2009MNRAS.395.1905C}. These galaxies are likely the progenitors of massive early-type galaxies found at $z \sim 2$ \citep{2005ApJ...626..680D,2008ApJ...681L..53C}. Therefore, it is necessary to study the characteristics of such systems to get a comprehensive view of star formation from the earliest to the current epochs.

Following the analysis presented in \citealt{2018MNRAS.477.4380S} (hereafter, \citetalias{2018MNRAS.477.4380S}), we study the SFR in different regions of AzTEC-1, a non-lensed starburst galaxy at $z \approx 4.3$ discovered in the Cosmic Evolution Survey (COSMOS) field \citep{2007ApJS..172...38S} with the AzTEC camera \citep{2008MNRAS.386..807W} on the \textit{James Clarke Maxwell Telescope} \citep{2008MNRAS.385.2225S}. A follow-up survey by the Large Millimeter Telescope found its spectroscopic redshift to be $4.3420\pm0.0004$ \citep{2015MNRAS.454.3485Y}. With a total $\lambda_{\mathrm{obs}}=860\,\micron$ continuum flux of $\sim 17\,\mathrm{mJy}$ and dust luminosity exceeding $10^{13}\,\lsol$ (\citealt{2018Natur.560..613T}, hereafter, \citetalias{2018Natur.560..613T}), AzTEC-1 falls in the commonly used definition of submillimeter galaxies (SMGs, \citealt{2011ApJ...743..159H}). \citetalias{2018MNRAS.477.4380S} presented the first tests of different star formation relations on the spatially resolved star-forming nucleus of a high-redshift starburst galaxy. However, the necessity of excellent spatial resolution limited the analysis to the lensed source SDP 81. Now, with the $\sim 550\,\mathrm{pc}$ resolution data at $z \approx 4.3$ from the Atacama Large Millimeter/Submillimeter Array (ALMA), we can test these relations at an even higher redshift in an unlensed clumpy disc galaxy. Such an analysis can help us understand what factors power high SFRs in high-redshift starbursts.

Section \ref{s:data} summarizes the ALMA observations of the continuum emission (\citealt{2016ApJ...829L..10I}; \citetalias{2018Natur.560..613T}) and CO$\,$(4-3) transition \citepalias{2018Natur.560..613T} of AzTEC-1 that we use in our work. Section \ref{s:clump_props} describes the calculation of the parameters that go into the star formation relations that we are testing. Section \ref{s:tests} discusses the comparison of the SFR we observe in AzTEC-1 with that predicted from various star formation relations published in the literature. Finally, we summarize our findings in Section \ref{s:conclusions}. We adopt the $\Lambda$CDM cosmology with H$_0 = 70\,\mathrm{km\,s}^{-1}\,\mathrm{Mpc}^{-1}$, $\Omega_\mathrm{m}$ = 0.27, $\Omega_{\Lambda}$ = 1-$\Omega_\mathrm{m}$ \citep{2003ApJS..148..175S} and the Chabrier IMF \citep{2003ApJ...586L.133C}. The luminosity distance and scale length corresponding to these parameters is $39.5\,\mathrm{Gpc}$ and $6.71\,\kpc\,/\arcsec$, respectively, for $z = 4.342$ \citep{2006PASP..118.1711W}.

\begin{figure}
\includegraphics[width=1.0\linewidth]{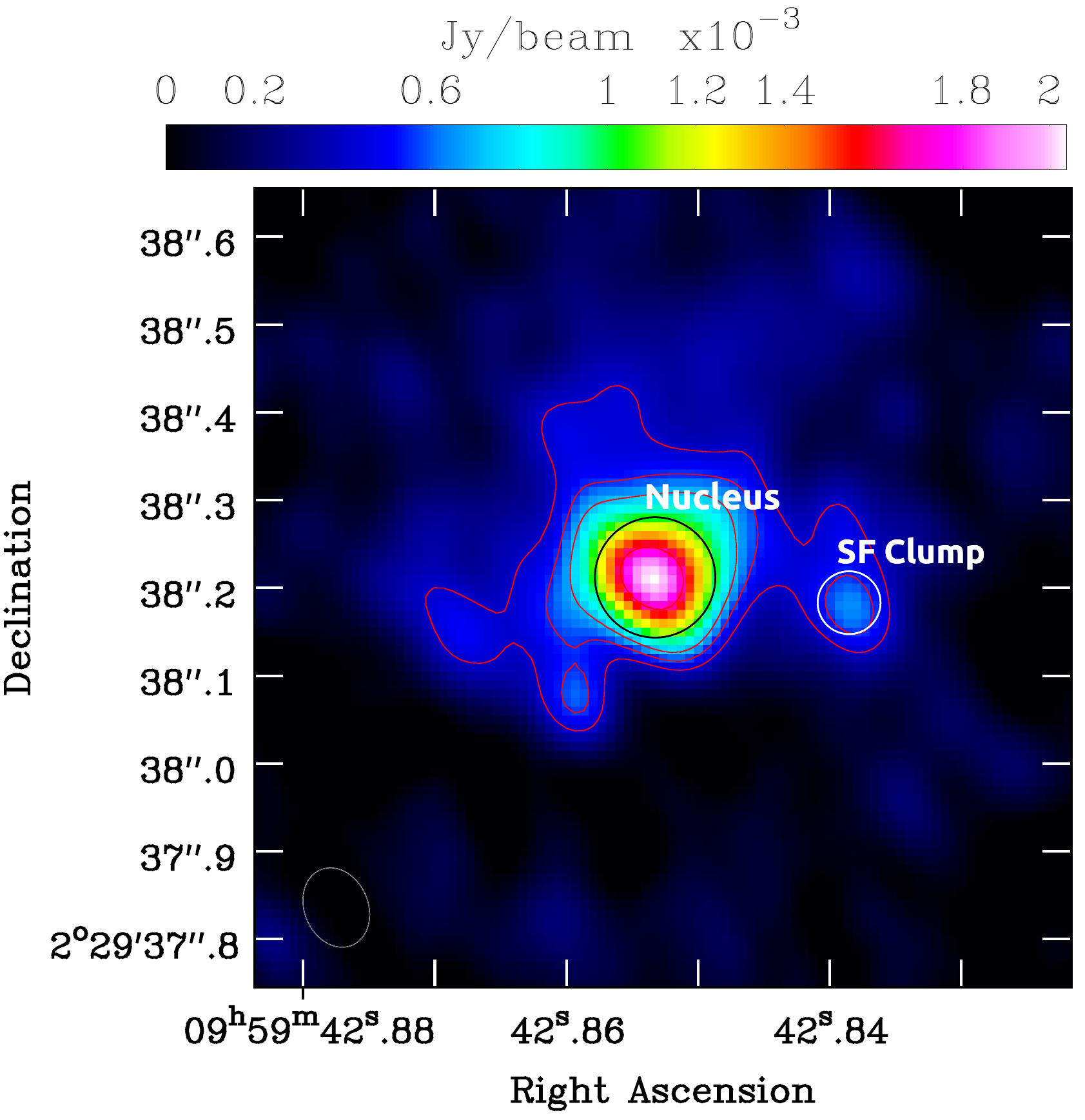}
\includegraphics[width=1.0\linewidth]{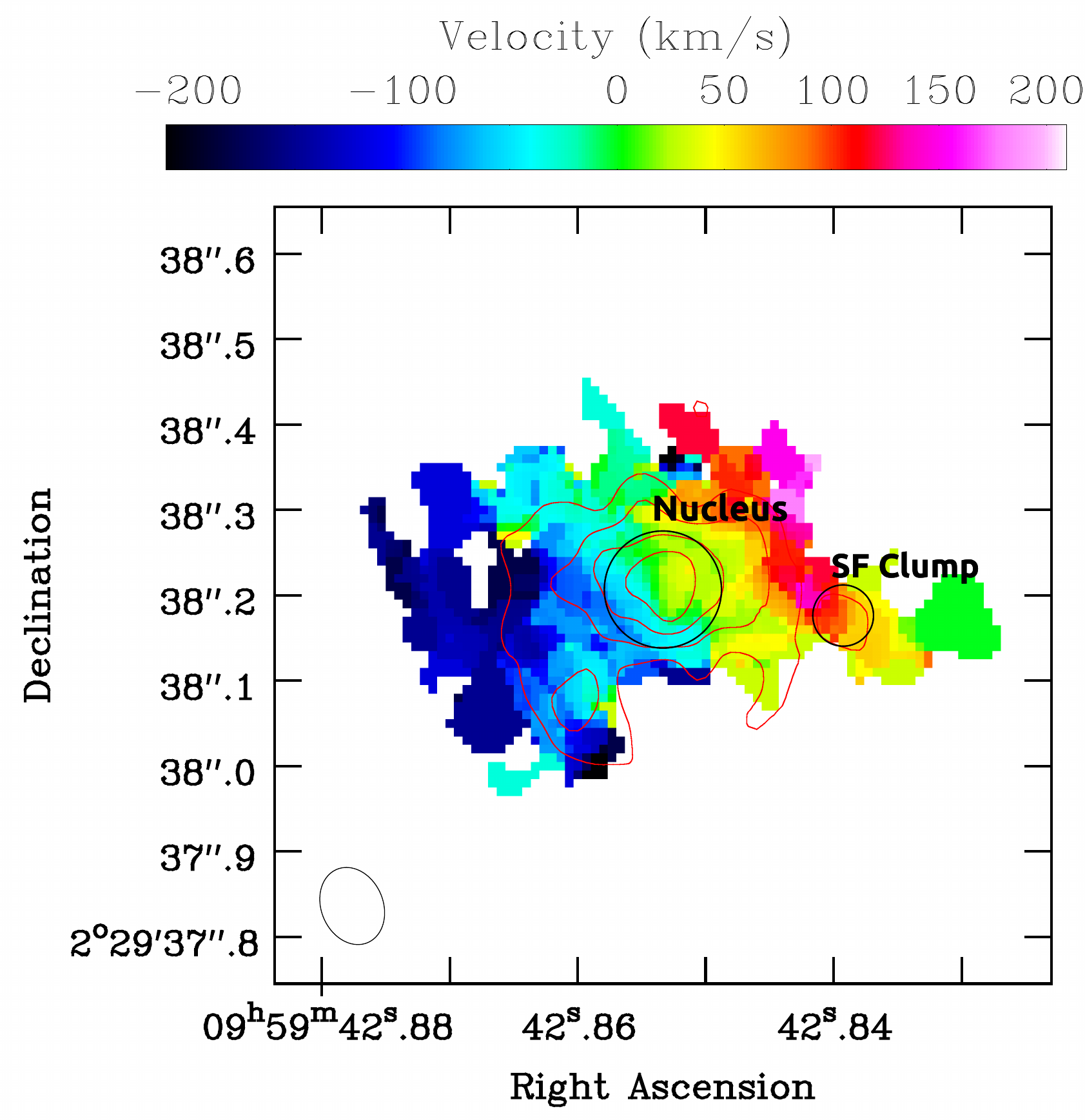}
\caption{$0.093''\times 0.072''$ ($624 \times 483\,\mathrm{pc}$) resolution maps of $\lambda_{\mathrm{obs}} = 860\,\micron$ continuum flux density and CO$\,$(4-3) velocity in AzTEC-1 (beam size as depicted in the lower left corners). The white and black circles in the two panels depict the nucleus and the star-forming (SF) clump we analyze in this work. The red contours in the continuum map correspond to 5, 7, 9 and $21\,\sigma_{860\,\micron}$ while those on the velocity map correspond to CO$\,$(4-3) velocity integrated flux density, plotted at 5, 7, 9 and $11\,\sigma_{\mathrm{CO}\,(4-3)}$.}
\label{fig:figure1}
\end{figure}

\section{Observations}
\label{s:data}
ALMA observations of the $\lambda_{\mathrm{obs}}=860\,\micron$ (Band 7) continuum emission in AzTEC-1 (centered at RA = $09^\mathrm{h}59^\mathrm{m}42.85^\mathrm{s}$, Dec = $+02^{\circ}29'38.23''$)  were carried out in November 2015 \citep{2016ApJ...829L..10I}. The $\lambda_{\mathrm{obs}}=3.2\,\mathrm{mm}$ (Band 3) continuum flux and CO$\,$(4-3) data were procured between October and November 2017 (\citetalias{2018Natur.560..613T}). The observations and data reduction are described in detail in the respective articles. The angular resolution of the data is $0.093\arcsec \times 0.072\arcsec$, corresponding to $624 \times 483\,\mathrm{pc}$ at $z \approx 4.3$ \citepalias{2018Natur.560..613T}. The total $860\,\micron$, $3.2\,\mathrm{mm}$ fluxes and the CO$\,$(4-3) velocity integrated flux measured by \citetalias{2018Natur.560..613T} are $S^{\mathrm{tot}}_{\,\nu\,,860\,\micron} = 17\pm1\,\mathrm{mJy}$, $S^{\mathrm{tot}}_{\,\nu\,,3.2\,\mathrm{mm}} = 273\pm41\,\mu\mathrm{Jy}$ and $S^{\mathrm{tot}}_{\mathrm{CO}}\,dv = 1.8\pm0.2\,\mathrm{Jy\,km\,s^{-1}}$, respectively.

Figure \ref{fig:figure1} shows the $\lambda_{\mathrm{obs}}=860\,\micron$ continuum map and the CO$\,$(4-3) velocity structure of the galaxy. The continuum map shows the compact structure of the nucleus of the galaxy, with a few outlying clumps. The velocity map clearly shows a large-scale gradient in the galaxy probably owing to its rotational motion, as has been observed for numerous other high-redshift sources (e.g., \citealt{2015ApJ...806L..17S,2017ApJ...841L..25T,2018MNRAS.476.3956T,2018Natur.553..178S}). We model the starburst nucleus as a Gaussian and define its diameter to be the full width at half maximum (FWHM) of the resulting Gaussian fit of the $860\,\micron$ continuum map\footnote{We also use the CO$\,$(4-3) velocity integrated flux map to perform the fit and find an agreement with the $860\,\micron$ continuum map for the sizes of the nucleus and the SF clump to within 14\% and 21\% respectively.}. While the galaxy exhibits multi-clump morphology, we find that a Gaussian fit local to the regions of interest (where the emission is peaked in the center) is a good approximation for the flux distribution. The radius we obtain for the nucleus is $R = 0.46\pm0.05\,\mathrm{kpc}$. We also study one off-center star-forming (SF) clump at RA = $09^\mathrm{h}59^\mathrm{m}42.84^\mathrm{s}$, Dec = $+02^{\circ}29'38.18''$. Like the nucleus, we also model this clump as a Gaussian and the FWHM gives us a radius of $0.24\pm0.02\,\mathrm{kpc}$. The following analyses are restricted to the nucleus and the SF clump because they are spatially distinct, have sufficient resolution to conduct the kinematic analysis (as we discuss in Section \ref{s:mach}) and can be approximated to first order as spherical regions, to estimate their volume densities. Such an analysis can also inform us about the spatially diverse star formation history of the galaxy. 

\begin{figure}
\includegraphics[width=1.0\linewidth]{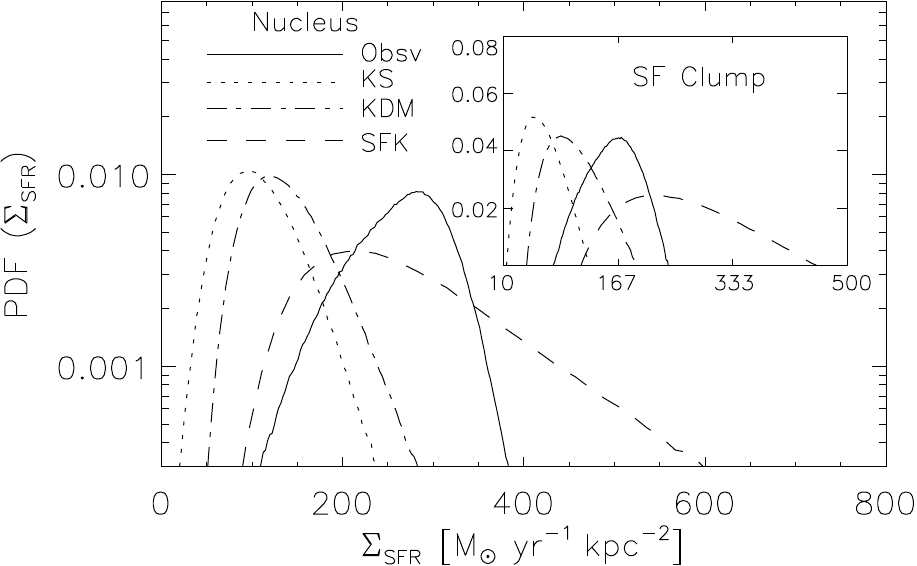}
\caption{Probability distribution function (PDF) of the star formation rate (SFR) surface density ($\sigmasfr$) in the nucleus, estimated from the data and those predicted from \citetalias{1998ApJ...498..541K} (dotted curve, equation \ref{eq:kslaw}), \citetalias{2012ApJ...745...69K} (dot-dashed curve, equation \ref{eq:kdmlaw}) and \citetalias{2015ApJ...806L..36S} (dashed curve, equation \ref{eq:sfklaw}) relations. Inset depicts the same PDFs for the star-forming (SF) clump.}
\label{fig:pdf_allsfr}
\end{figure}

\section{Physical Parameters for the Nucleus and SF Clump}
\label{s:clump_props}
In this Section, we measure the parameters that go into the three star formation relations we test in Section \ref{s:tests}. We use a Monte Carlo analysis to estimate and propagate the uncertainties on all the parameters. We summarize the analysis in subsequent subsections, present the calculated quantities in Table \ref{tab:table1} and refer the reader to \citetalias{2018MNRAS.477.4380S} for details on the procedure.

\begin{table*}
\centering
\caption{Properties of the two regions we study (nucleus and star-forming (SF) clump) in AzTEC-1 with the mean, and the standard deviation quoted in brackets. For the SFR surface densities ($\sigmasfr$), we tabulate the median (standard deviation quoted in brackets) of their PDFs. We also report the difference between the median and the $16^{\mathrm{th}}$ and $84^{\mathrm{th}}$ percentiles of the different parameters (as subscripts and superscripts, respectively).}
\begin{tabular}{lcrr}
\hline
Parameter & Symbol/Unit & Nucleus & SF Clump\\
\hline
Radius & $R$/$\mathrm{kpc}$ & $0.46\,(0.05)$ & $0.24\,(0.02)$\\
Area & $A$/$\mathrm{kpc}^2$ & $0.65\,(0.13)$ & $0.18\,(0.04)$\\
$860\,\micron$ Flux & $S_{\,\nu,860\,\micron}$/$\mathrm{mJy}$ & $2.82\,(0.11)$ & $0.48\,(0.02)$\\
$3.2\,\mathrm{mm}$ Flux & $S_{\,\nu,3.2\,\mathrm{mm}}$/$\mu\mathrm{Jy}$ & $59\,(7)$ & $14\,(4)$\\
CO$\,$(4-3) Flux & $S_{\mathrm{\,\nu,CO}\,(4-3)}dv$/$\mathrm{Jy}\,\mathrm{km\,s}^{-1}$ & $0.24\,(0.02)$ & $0.052\,(0.004)$\\
Velocity Dispersion & $\sigma_{v,\textrm{turb}}$/$\mathrm{km\,s}^{-1}$ & $12\,(1)$ & $17\,(2)$\\
Mach Number & $\mach$ & $35\,(16)$ & $50\,(20)$\\
Gas Mass & $M_{\mathrm{gas}}$/$10^9\,\msol$ & $9.9\,(2.7)^{+2.8}_{-0.9}$ & $1.9\,(0.5)^{+0.9}_{-0.4}$\\[0.07cm]
Gas Surface Density & $\Sigma_{\textrm{gas}}$/$10^{10}\,\msol\,\mathrm{kpc}^{-2}$ & $1.5\,(0.3)^{+1.7}_{-0.5}$ & $1.0\,(0.2)^{+0.8}_{-0.3}$\\[0.07cm]
Gas Volume Density & $\rho$/$10^{-21}\,\mathrm{g\,cm^{-3}}$ & $1.8\,(0.8)^{+1.3}_{-0.5}$ & $2.4\,(1.0)^{+1.7}_{-0.6}$\\[0.07cm]
Freefall Time & $\tff$/$\mathrm{Myr}$ & $1.7\,(0.5)^{+0.9}_{-0.4}$ & $1.5\,(0.3)^{+0.8}_{-0.4}$\\[0.07cm]
\hline
Measured SFR & $\Sigma_{\mathrm{SFR}}$/$\,\msol\,\yr^{-1}\,\kpc^{-2}$ & $270\,(54)^{+74}_{-105}$ & $170\,(38)^{+26}_{-39}$\\
\hline
Predicted SFRs & $\Sigma_{\mathrm{SFR,KS}}$/$\,\msol\,\yr^{-1}\,\kpc^{-2}$ & $105\,(42)^{+170}_{-75}$ & $62\,(26)^{+100}_{-20}$\\[0.07cm]
&$\Sigma_{\mathrm{SFR,KDM}}$/$\,\msol\,\yr^{-1}\,\kpc^{-2}$ & $134\,(47)^{+240}_{-55}$ & $106\,(40)^{+190}_{-65}$\\[0.07cm]
&$\Sigma_{\mathrm{SFR,SFK}}$/$\,\msol\,\yr^{-1}\,\kpc^{-2}$ & $270\,(145)^{+520}_{-120}$ & $280\,(147)^{+500}_{-115}$\\
\hline
\end{tabular}
\label{tab:table1}
\end{table*}

\subsection{Star Formation Rate}
\label{s:sfr}
We follow \citetalias{2018Natur.560..613T} to estimate the SFR per unit area in the two regions as $\sigmasfr = \mathrm{SFR^{tot}} \times (S_{\,\nu,\,860\,\micron}/S^{\mathrm{tot}}_{\,\nu,\,860\,\micron})/A$, where $A$ is the effective area of the region that we find from the 2D Gaussian fit in Section \ref{s:data}, and $\mathrm{SFR^{tot}}$ is the total SFR of the galaxy. By fitting the spectral energy distribution (SED) of the galaxy at multiple wavelengths, \citetalias{2018Natur.560..613T} find $\mathrm{SFR^{tot}}=1186^{+36}_{-291}\,\msol\,\mathrm{yr^{-1}}$. 

We integrate the area under the modeled Gaussian curve to estimate the $860\,\micron$ flux in the two regions. For the nucleus and the SF clump, we obtain $S_{\,\nu\,,860\,\micron} = 2.82\pm0.11\,\mathrm{mJy}$ and $0.48\pm0.02\,\mathrm{mJy}$, respectively. We plot the probability density function (PDF) of the measured SFR surface densities ($\Sigma_{\mathrm{SFR}}$) for the two regions in Figure \ref{fig:pdf_allsfr} (solid lines). The PDFs give the median SFR per unit area: $\Sigma^{\mathrm{nucleus}}_{\mathrm{SFR}} = 270\pm54\,\sfrunits$ and $\Sigma^{\mathrm{sfclump}}_{\mathrm{SFR}} = 170\pm38\,\sfrunits$ for the nucleus and the SF clump, respectively. SFR surface densities of similar magnitudes have been found in numerous other high-redshift starbursts (\citealt{2015ApJ...798L..18H,2015ApJ...810..133I,2018MNRAS.475.3467E}; \citetalias{2018MNRAS.477.4380S}). 

\begin{figure*}
\includegraphics[width=1.0\linewidth]{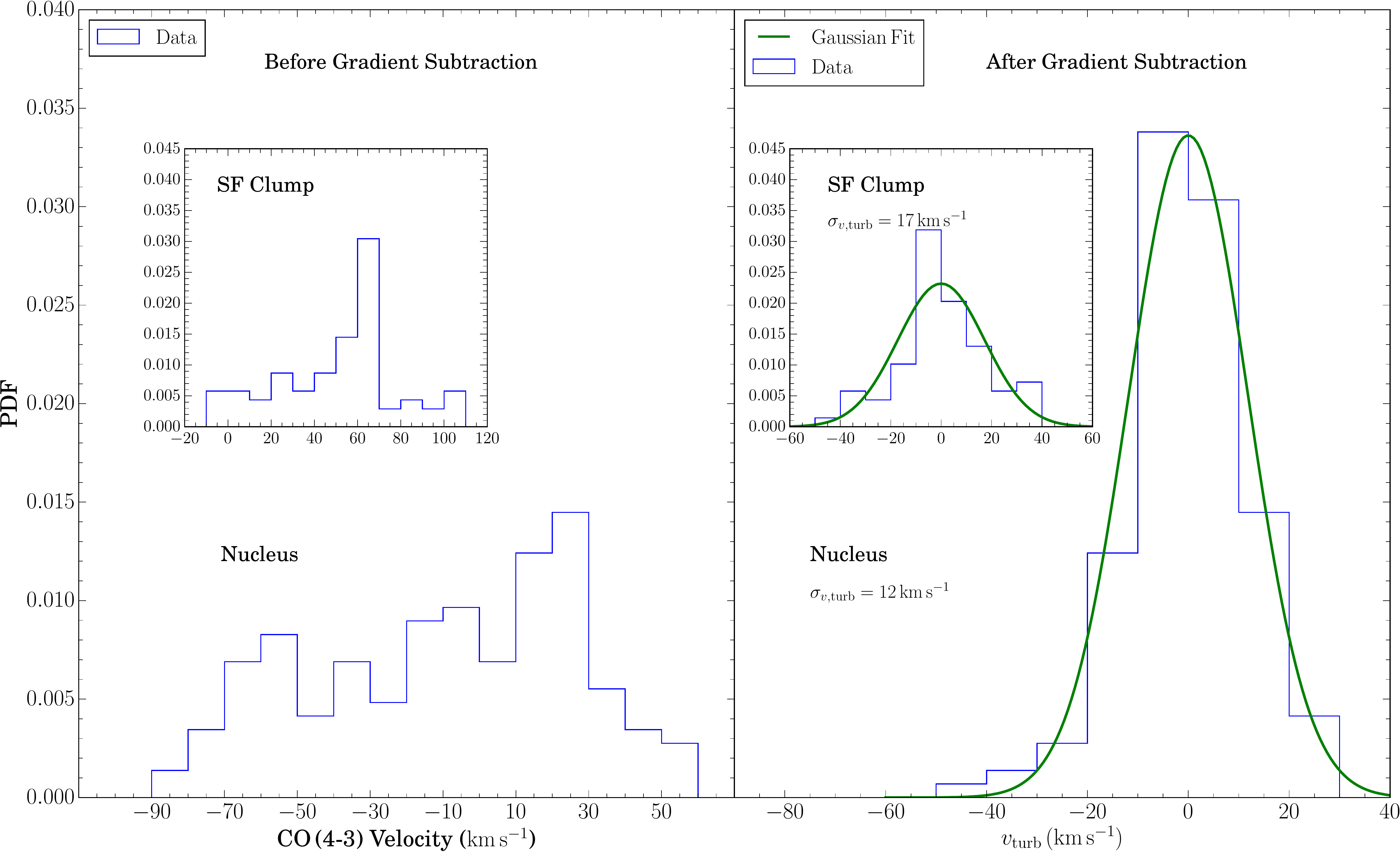}
\caption{PDF of the CO$\,$(4-3) velocities (as labelled in Figure \ref{fig:figure1}) and velocities after the subtraction of the local, linear velocity gradient ($v_\mathrm{turb}$). The standard deviation ($\sigma_{v,\mathrm{turb}}$) of the turbulent velocity (gas velocity after gradient subtraction) yields the turbulent Mach number ($\mach$) in the two regions. Inset depicts the same for the star-forming (SF) clump.}
\label{fig:pdf_turbvels}
\end{figure*}

\subsection{Velocity Dispersion and Mach Number}
\label{s:mach}
As we see from Figure \ref{fig:figure1}, the CO$\,$(4-3) velocity map shows a large-scale velocity gradient in the galaxy due to its rotational motion. To extract the turbulent velocity features in the regions, we follow the analysis presented in Section 4 of \citetalias{2018MNRAS.477.4380S}, \textit{i.e.,} we fit and subtract the local, linear velocity gradient in the regions (see also, \citealt{2016ApJ...832..143F}). This allows us to subtract the contributions of rotation to the dispersion and hence yields an estimate of the turbulent dispersion, provided there is enough spatial resolution. We use the resolution check algorithm described in Section 4.1 of \citetalias{2018MNRAS.477.4380S} and find that we have sufficient resolution in the regions, which is necessary for the convergence of the derived turbulent velocity dispersion.

The turbulent velocity dispersion we obtain through the local, linear velocity gradient fit method in the nucleus and the SF clump are $\sigma_{v,\mathrm{turb}} = 12\pm1\,\mathrm{km\,s^{-1}}$ and $17\pm2\,\mathrm{km\,s^{-1}}$, respectively. We show the PDFs of the velocities in the regions before and after the subtraction of this gradient in Figure \ref{fig:pdf_turbvels}. The gradient-subtracted velocity PDF is consistent with a Gaussian, like those estimated in other star-forming regions (\citealt{2016ApJ...832..143F}; \citetalias{2018MNRAS.477.4380S}) and predicted by simulations of supersonic turbulence \citep{2000ApJ...535..869K,2013MNRAS.436.1245F}. 

To independently check the validity of the linear gradient fit algorithm and given that the disc of AzTEC-1 is rotationally supported \citepalias{2018Natur.560..613T}, we also fit the CO datacube with a rotating disc model to account for the large-scale rotation of the galaxy and correct for beam smearing. For this purpose, we use the code $^{\mathrm{3D}}$BAROLO \citep{2015MNRAS.451.3021D} which fits tilted-ring models \citep{1974ApJ...193..309R} to spectroscopic datacubes and is applicable to multi-wavelength observations \citep{2016ApJ...823...68S,2018MNRAS.476.5417S}. We present the model and the residual maps in Appendix \ref{s:appendix}. The velocity dispersions we obtain for the nucleus and the SF clump are $13$ and $23\,\mathrm{km\,s^{-1}}$ respectively. From Figure \ref{fig:appendix}, we notice that the residual velocities in the nucleus do not show any residual gradient after subtraction (implying that the nucleus follows the galaxy-wide rotation) and is consistent with the velocity dispersion we obtain from the local, linear velocity gradient fit. However, we notice a leftover gradient in the SF clump which, when accounted for using a linear gradient fit, gives a velocity dispersion of $18\,\mathrm{km\,s^{-1}}$, in excellent agreement with what we obtain from the former method. This implies that the SF clump has some intrinsic rotation of its own which is different than the systematic galaxy-wide rotation which should be subtracted to reveal the turbulent features.

Assuming the temperature of the molecular gas in the two regions to be between 10-100 K, we find the sound speed as $c_{\mathrm{s}} = 0.4\pm0.2\,\mathrm{km\,s^{-1}}$ (e.g., \citealt{2016ApJ...832..143F}), where the error represents the range of gas temperatures we consider. Then, the turbulent Mach number is given by: $\mach = \sigma_{v,\mathrm{turb}}/c_{\mathrm{s}}$. The Mach numbers we obtain for the nucleus and the SF clump are $35\pm16$ and $50\pm20$, respectively. These Mach numbers are of the same order of magnitude as the few predicted for starburst environments at low and high redshifts (\citealt{2015ApJ...806L..36S,2017MNRAS.468.3965F}; \citetalias{2018MNRAS.477.4380S}).

\subsection{Molecular Gas Mass}
\label{s:gasmass_and_tff}
CO is often used as a tracer for the cold and dense molecular gas present in star-forming regions because it is bright, easily observable due to its dipole moment and the second-most abundant molecule in star-forming regions \citep{2015A&A...577A..46D,2018A&ARv..26....5C}. Following \citetalias{2018Natur.560..613T}, we measure the molecular gas mass per unit area in the regions ($\siggas$) as: $\Sigma_{\mathrm{gas}} = M^{\mathrm{tot}}_{\mathrm{gas}} \times (S_{\mathrm{CO}}dv/S^{\mathrm{tot}}_{\mathrm{CO}}dv)/A$, where $S^{\mathrm{tot}}_{\mathrm{CO}}\,dv$ is the total CO$\,$(4-3) velocity-integrated flux density in the galaxy. The CO$\,$(4-3) velocity-integrated flux densities we find in the nucleus and the SF clump are $0.24\pm0.02\,\mathrm{Jy\,km\,s^{-1}}$ and $0.052\pm0.004\,\mathrm{Jy\,km\,s^{-1}}$ respectively. We convert them to CO$\,$(4-3) line luminosities \citep{2005ARA&A..43..677S} and scale them to CO$\,$(1-0) line luminosities with the CO excitation scaling factor $r_{43} = L'_{\mathrm{CO\,(4-3)}}/L'_{\mathrm{CO\,(1-0)}}$. We follow \citetalias{2018Natur.560..613T} who set $r_{43}=0.91$ to ensure consistency between CO and C$\,$I gas masses, however, we also experiment with $r_{43}=0.46$, which is the average value for SMGs (\citealt{2013ARA&A..51..105C}, see also \citealt{2011MNRAS.412..287N}). This is a significant systematic that can change the derived SFRs by a factor of $\sim 2-3$ and we include it in our error propagation. We transform CO$\,$(1-0) line luminosity to gas (H$_2$) mass using a CO-to-H$_2$ conversion factor $\alpha_{\mathrm{CO}} = 0.8\pm0.1\,\msol\,\mathrm{K^{-1}}\,\mathrm{km^{-1}\,s}\,\pc^{-2}$ \citep{2013ARA&A..51..207B,2013ARA&A..51..105C}. 

Putting these parameters together we find $M^{\mathrm{tot}}_{\mathrm{gas}} = (7.4\pm1.1)\times 10^{10}\,\msol$. The molecular gas masses we estimate for the two regions are $M_{\mathrm{gas}} = (9.9\pm2.7)\times 10^{9}\,\msol$ and $(1.9\pm0.5)\times 10^{9}\,\msol$, respectively. We also estimate the gas masses from the dust masses that can be obtained from the Rayleigh-Jeans (RJ) tail of the SED \citep{2012ApJ...760....6M,2014ApJ...783...84S,2016ApJ...820...83S}, by assuming a gas-to-dust ratio of 100. The gas mass we get for the nucleus is $M_{\mathrm{gas,RJ}} = (1.4\pm0.4)\times10^{10}\,\msol$, in good agreement with that found using the CO$\,$(4-3) data. For the SF clump, we get $M_{\mathrm{gas,RJ}} = (3.0\pm0.9)\times10^{9}\,\msol$ which is consistent with the CO based gas mass within the systematic uncertainty. 

Further, we calculate the gas surface density for the two regions as $\Sigma_{\mathrm{gas}} = (1.5\pm0.3)\times 10^{10}\,\msol\,\mathrm{kpc^{-2}}$ and $(1.0\pm0.2)\times10^{10}\,\msol\,\mathrm{kpc^{-2}}$, respectively. It is interesting to note that $\Sigma_{\mathrm{gas}}$ of the nucleus of AzTEC-1 is almost twice that of the nuclear region of SDP~81, but the rate of collapse of the gas is similar (as we show in Section \ref{s:kdmrel}). Table \ref{tab:table1} summarizes all measured and derived parameters for the nucleus and the SF clump.

\section{SFR Predicted by Different Star Formation Relations}
\label{s:tests}

\subsection{Kennicutt-Schmidt (KS) Relation}
\label{s:ksrel}
Firstly, we test the Kennicutt-Schmidt (\citetalias{1998ApJ...498..541K}, \citealt{1998ApJ...498..541K}) relation, which connects $\Sigma_{\mathrm{gas}}$ of a star-forming region to its SFR surface density ($\Sigma_{\mathrm{SFR}}$) via a power-law
\begin{equation}
\centering
\label{eq:kslaw}
\Sigma_{\textrm{SFR,KS}} = (1.6\pm0.4)\times10^{-4}\, \bigg(\frac{\Sigma_{\mathrm{gas}}}{\msol\,\mathrm{pc^{-2}}}\bigg)^N\,\sfrunits,
\end{equation}
where $N=1.40\pm0.15$ was empirically derived by fitting the gas surface density against the SFR surface density, and the constant has been corrected for the Chabrier IMF \citep{2008ApJ...680..246T,2010MNRAS.403.1894D}. We show the PDF of the $\Sigma_{\mathrm{SFR}}$ predicted by the \citetalias{1998ApJ...498..541K} relation as dotted lines in Figure \ref{fig:pdf_allsfr}. The median of the PDFs give $\Sigma^{\mathrm{nucleus}}_{\mathrm{SFR,KS}} = 105\pm42$ and $\Sigma^{\mathrm{sfclump}}_{\mathrm{SFR,KS}} = 62\pm26\,\sfrunits$. The \citetalias{1998ApJ...498..541K} relation underestimates the $\Sigma_{\mathrm{SFR}}$ in both regions by a factor 2-3. However, when the systematic uncertainty on $r_{43}$ is included, the KS relation can explain the measured $\Sigma_{\mathrm{SFR}}$ to within $2\sigma$ for the nucleus.

\begin{figure*}
\includegraphics[width=1.0\linewidth, angle=0]{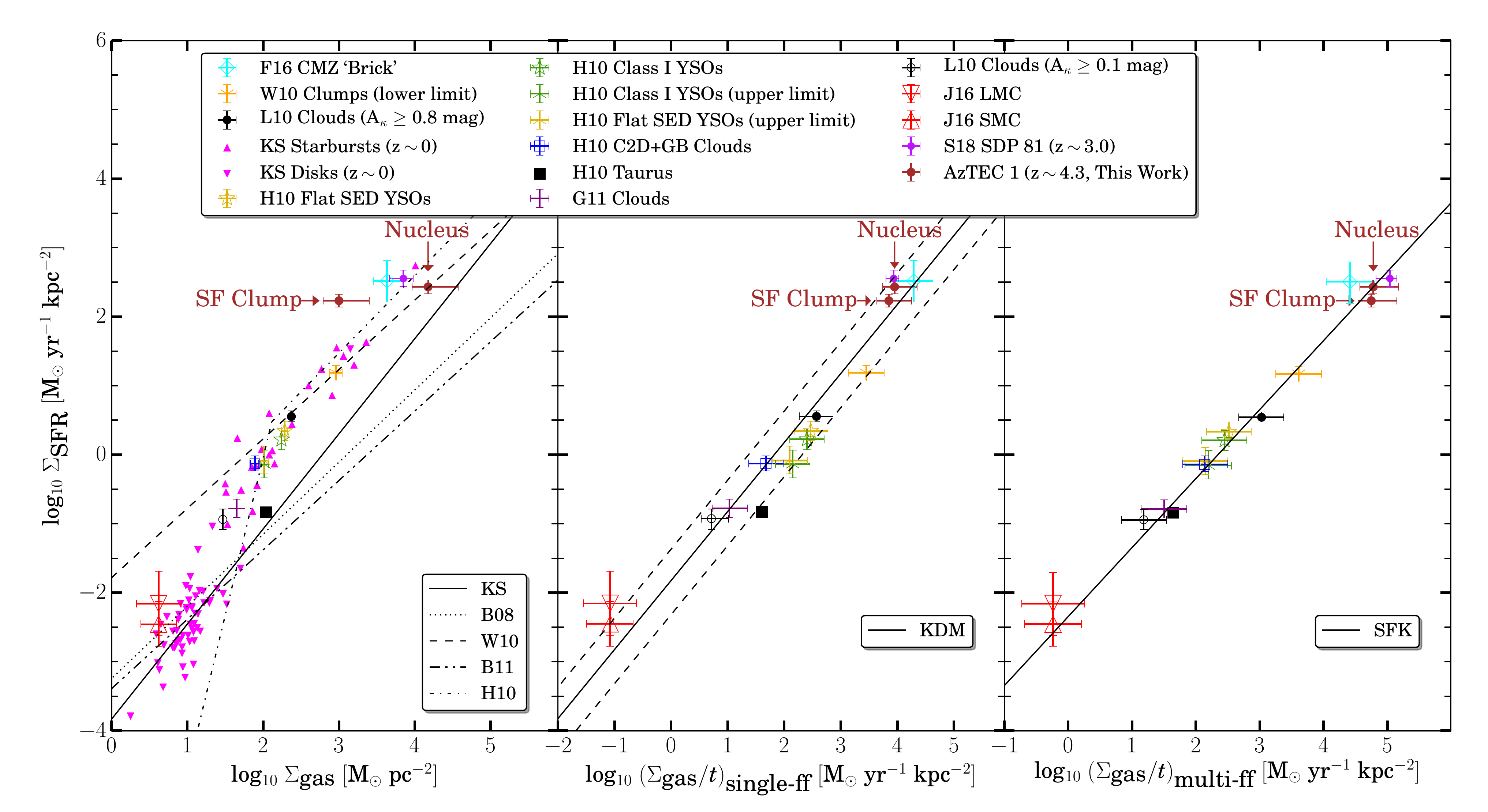}
\caption{Comparison of star formation relations in multiple datasets, showing the measured and predicted SFRs, overlaid with the two regions of AzTEC-1 in brown. Asymmetric error bars for AzTEC-1 regions are a result of the systematic uncertainty on the CO excitation scale factor $r_{43}$. \textit{Left-panel:} Measured SFR surface density as a function of the gas surface density (equation \ref{eq:kslaw}) in numerous star-forming regions (\protect\citealt{2010ApJ...723.1019H} (H10), \citealt{2010ApJ...724..687L} (L10), \citealt{2010ApJS..188..313W} (W10), \citealt{2011ApJ...739...84G} (G11), \citealt{2016ApJ...825...12J} (J16), \citealt{2016ApJ...832..143F} (F16) and \citealt{2018MNRAS.477.4380S} \citepalias{2018MNRAS.477.4380S}). Also plotted are the SFR relations proposed by \citealt{2008AJ....136.2846B} (B08), \citealt{2011ApJ...730L..13B} (B11), \citealt{2010ApJS..188..313W} (W10) and \citealt{2010ApJ...723.1019H} (H10). \textit{Middle-panel:} Measured SFR surface density plotted against the single-freefall time model by \protect\citealt{2012ApJ...745...69K} (equation \ref{eq:kdmlaw}). The dashed lines depict deviations by a factor of 3 from the best-fit relation \citep[see also][]{2013ApJ...779...89K}. \textit{Right-panel:} Measured SFR surface density plotted against the multi-freefall model of \citealt{2015ApJ...806L..36S} (equation \ref{eq:sfklaw}).}
\label{fig:all_laws_figure}
\end{figure*}

\subsection{Krumholz-Dekel-McKee (KDM) Relation}
\label{s:kdmrel}
We move on to the single-freefall model given by \citealt{2012ApJ...745...69K} (\citetalias{2012ApJ...745...69K}) which also takes into account the freefall time of the gas under collapse in its prediction of $\Sigma_{\mathrm{SFR}}$
\begin{equation}
\label{eq:kdmlaw}
\Sigma_{\textrm{SFR,KDM}} = f_{\textrm{H}_{2}} \epsilon_{\textrm{ff}} \frac{\Sigma_{\textrm{gas}}}{t_{\textrm{ff}}}\,,
\end{equation}
where $f_{\textrm{H}_{2}}$ is a factor of order unity and $\epsilon_{\textrm{ff}}$ is the star formation efficiency, found to be $0.015$ \citep{2013ApJ...779...89K}. $t_{\mathrm{ff}}$ is the freefall timescale of collapse, given by $t_{\mathrm{ff}} = \sqrt{3\pi/32G\rho}$, where $\rho$ is the volume density of the region. Following \citetalias{2012ApJ...745...69K}, we approximate the two regions as regular spheres, and find the volume densities in the nucleus and the SF clump to be $\rho = (1.8\pm0.8) \times 10^{-21}\,\mathrm{g\,cm^{-3}}$ and $(2.4\pm1.0)\times 10^{-21}\,\mathrm{g\,cm^{-3}}$, respectively. Using these, the freefall time we obtain is $t_{\mathrm{ff}} = 1.7\pm0.5\,\mathrm{Myr}$ and $1.5\pm0.3\,\mathrm{Myr}$, respectively. The median SFR surface densities we obtain from the \citetalias{2012ApJ...745...69K} relation are $\Sigma^{\mathrm{nucleus}}_{\mathrm{SFR,KDM}} = 134\pm47$ and $\Sigma^{\mathrm{sfclump}}_{\mathrm{SFR,KDM}} = 106\pm40\,\sfrunits$. We plot their PDFs in Figure \ref{fig:pdf_allsfr} (dash-dot lines). Although the \citetalias{2012ApJ...745...69K} relation underestimates the median $\Sigma_{\mathrm{SFR}}$ in the nucleus and the SF clump by a factor $\sim$ 2 and 1.6 respectively, its predictions are consistent with the measured values when the systematic uncertainties are included. Further refinement may be possible as we move towards a larger sample.

\subsection{Salim-Federrath-Kewley (SFK) Relation}
\label{s:sfkrel}
\citealt{2015ApJ...806L..36S} (\citetalias{2015ApJ...806L..36S}) extended the \citetalias{2012ApJ...745...69K} relation to include the effects of physical variations of turbulence and magnetic field strength on star formation. Their multi-freefall model takes the form
\begin{equation}
\centering
\label{eq:sfklaw}
\Sigma_{\mathrm{SFR,SFK}} = \epsilon_{\mathrm{ff}}\,\frac{\Sigma_{\mathrm{gas}}}{t_{\mathrm{ff}}} \,\Big[{1+b^2\mach^2\frac{\plasmabeta}{\plasmabeta+1}}\Big]^{3/8}\,,
\end{equation}
where $\epsilon_{\mathrm{ff}}=0.0045$, $b$ is the turbulent driving parameter (set to 0.4 to reflect a mixed turbulent driving mode, see \citealt{2010A&A...512A..81F,2012ApJ...761..156F}) and $\plasmabeta$ is the ratio of the thermal to magnetic pressure \citep{2012MNRAS.423.2680M}. We lack the magnetic field strength measurements for AzTEC-1; following \citetalias{2018MNRAS.477.4380S}, we set $\plasmabeta \to \infty$ such that $\plasmabeta/(\plasmabeta+1) = 1$. This means we assume that the magnetic field is zero.

The median of the PDF of $\Sigma_{\mathrm{SFR}}$ from the multi-freefall \citetalias{2015ApJ...806L..36S} model is $\Sigma^{\mathrm{nucleus}}_{\mathrm{SFR,SFK}} = 270\pm145$ and $\Sigma^{\mathrm{sfclump}}_{\mathrm{SFR,SFK}} = 280\pm147\,\sfrunits$. Figure \ref{fig:pdf_allsfr} (dashed lines) shows that the $\Sigma_{\mathrm{SFR}}$ we obtain from the \citetalias{2015ApJ...806L..36S} relation agrees with the measured $\Sigma_{\mathrm{SFR}}$ within the uncertainties. The slight overestimation of the $\Sigma_{\mathrm{SFR}}$ in the SF clump may be the result of ignoring magnetic fields in our calculations. It has been shown that finite, typical magnetic field strengths (for Milky Way conditions) can reduce SFRs by up to a factor of 2-3 \citep{2011ApJ...730...40P,2012ApJ...761..156F,2015MNRAS.450.4035F}. 

\subsection{Comparison of Star Formation Relations Across Multiple Datasets}
\label{s:sfrcompares}
We compare the measured SFRs in local star-forming regions with the high-redshift starburst galaxy SDP~81 ($z \approx 3.0$, \citetalias{2018MNRAS.477.4380S}) and AzTEC-1 ($z \approx 4.3$) against the three star formation relations we discussed. We present the comparison in Figure \ref{fig:all_laws_figure}, which has been adapted from \citetalias{2018MNRAS.477.4380S}. We also incorporate some other star formation relations based on gas surface densities in the left panel of Figure \ref{fig:all_laws_figure}. These star formation relations do not predict reasonable SFRs for a large subset of the diverse sample. 

When the freefall time of the gas is included (the \citetalias{2012ApJ...745...69K} relation), some of the scatter in the measured SFR can be accounted for, as we show in the middle panel of Figure \ref{fig:all_laws_figure}. The remaining scatter is substantially lower in the \citetalias{2015ApJ...806L..36S} relation (right panel of Figure \ref{fig:all_laws_figure}) compared to the \citetalias{1998ApJ...498..541K} and \citetalias{2012ApJ...745...69K} relations. This is because the \citetalias{2015ApJ...806L..36S} relation does not only take gravity into account (as \citetalias{2012ApJ...745...69K} does), but it also adds the physical effects of turbulence and magnetic fields, which are crucial for star formation \citep{2005ApJ...630..250K,2011ApJ...730...40P,2011ApJ...743L..29H,2012ApJ...761..156F,2015MNRAS.450.4035F}. However, while the \citetalias{2015ApJ...806L..36S} relation has the physics of magnetic fields included, we cannot make use of that feature because we do not know the magnetic field strengths in high-redshift galaxies. Nevertheless, we find that the \citetalias{2015ApJ...806L..36S} relation best predicts the SFRs for both low- and high-redshift regions. Given the large systematic uncertainties that go in calculating the key ingredients, we require a large sample of diverse star-forming regions on multiple scales to fully assess the validity of these relations.

\section{Conclusions}
\label{s:conclusions}
In this work, we probe the star formation characteristics of the starburst galaxy AzTEC-1, at redshift $z \approx\,4.3$ in the COSMOS field. AzTEC-1 is one of the best resolved non-lensed galaxies at $z > 4$, and is interestingly forming stars at a rate $>\,1000\,\msol\,\mathrm{yr^{-1}}$. It has a suitable environment to study the characteristics of star formation and understand how high-redshift ($z > 4$) galaxies can sustain such high SFRs. Following the methodology described in \cite{2018MNRAS.477.4380S}, we use spatially resolved (sub-kiloparsec scale) ALMA observations of the sub-millimeter continuum and CO$\,$(4-3) emission to test the validity of three star formation relations in the literature. In particular, we study the galaxy nucleus and an off-center star-forming (SF) clump in this galaxy because they have sufficient resolution to apply the kinematic analysis.

The nucleus of AzTEC-1 has a very compact structure, with a gas surface density ($\Sigma_{\mathrm{gas}}$) $2\times$ the nucleus of the starburst galaxy SDP 81 at $z \approx 3.0$. However, its median SFR surface density ($\Sigma_{\mathrm{SFR}}$) is only 70\% of the latter, possibly because it is one-third as turbulent. Similarly, while the $\Sigma_{\mathrm{gas}}$ of the SF clump in AzTEC-1 is almost an order of magnitude lower than the galaxy nucleus in AzTEC-1, the $\Sigma_{\mathrm{SFR}}$ of the former is only two-fifth as high as the latter, possibly because it is $1.4\times$ more turbulent. While turbulence acts against star formation by stabilizing the cloud to prevent collapse on large scales, supersonic turbulence can create local shock-compressed regions which are the progenitors of star formation sites \citep{2011ApJ...740..120R,2011ApJ...743L..29H,2012ApJ...761..156F}. Thus, an interplay between gravity and turbulence seems to play a major role in sustaining high SFRs in these starbursts. We also find that the SF clump has an intrinsic rotation of its own, which does not follow the galaxy-wide rotation. Such a star-forming clump reflects the spatially diverse star formation history of the galaxy and adds valuable information about its past evolution.

We show that the Kennicutt-Schmidt \citepalias{1998ApJ...498..541K} relation underestimates the $\Sigma_{\mathrm{SFR}}$ in both the regions in AzTEC-1 by a factor of 2-3. The single-freefall model by \citealt{2012ApJ...745...69K} (\citetalias{2012ApJ...745...69K}) underestimates the median $\Sigma_{\mathrm{SFR}}$ in the nucleus and the SF clump by a factor $\sim$ 2 and 1.6 respectively, however, these predictions are within the systematic uncertainties. The multi-freefall model by \citealt{2015ApJ...806L..36S} (\citetalias{2015ApJ...806L..36S}) gives median $\Sigma_{\mathrm{SFR}}$ consistent with that measured in the nucleus and overpredicts it for the SF clump by 60\%. The slight overestimation from the \citepalias{2015ApJ...806L..36S} relation possibly arises because we lack the magnetic field strength in the galaxy, and neglect its effect on $\Sigma_{\mathrm{SFR}}$. AzTEC-1 thus forms a part of the very few sources (and the only source at $z \ga 2$, apart from SDP~81) for which the \citetalias{2012ApJ...745...69K} and \citetalias{2015ApJ...806L..36S} relations have been tested. 

Examining the performance of these relations across multiple datasets and given all the caveats, we conclude that the \citetalias{2015ApJ...806L..36S} relation provides the best prediction for the SFR in low- and high-redshift star-forming regions. We also find that an interplay between turbulence and gravity can help sustain high SFRs in high-redshift starburst galaxies. Our method can be used to reproduce the same analysis for other local and high-redshift star-forming regions on spatially-resolved scales, which can inform us about the diverse star formation history of these regions.

\section*{Acknowledgements}
We thank the anonymous referee for a constructive feedback which helped to improve the presentation. We thank Stephanie Monty and Harrison Abbot for useful discussions on flux densities. P.~S.~is supported by an Australian Government Research Training Program (RTP) Scholarship. E.~dC.~gratefully acknowledges the Australian Research Council for funding support as the recipient of a Future Fellowship (FT150100079). C.~F.~acknowledges funding provided by the Australian Research Council (Discovery Project DP170100603 and Future Fellowship FT180100495), and the Australia-Germany Joint Research Cooperation Scheme (UA-DAAD). E.~W.~acknowledges support by the Australian Research Council Centre of Excellence for All Sky Astrophysics in 3 Dimensions (ASTRO 3D), through project number CE170100013. E.~M.~D.~T.~acknowledges the support of the Australian Research Council through grant DP160100723. I.~A.~is supported through Consejo Nacional de Ciencia y Tecnolog\'ia, Mexico (CONACYT) grants FDC-2016-1848 and CB-2016-281948.

This paper uses data from ALMA programs ADS/JAO.ALMA \#2015.1.01345.S, \#2017.1.00300.S and \#2017.A.00032.S. ALMA is a partnership of ESO, NSF (USA), NINS (Japan), NRC (Canada), NSC and ASIAA (Taiwan), and KASI (Republic of Korea) and the Republic of Chile. The JAO is operated by ESO, AUI/NRAO and NAOJ. This research has also made use of NASA Astrophysics Data System.



\bibliographystyle{mnras}
\bibliography{main}

\appendix
\counterwithin{figure}{section}
\section{Kinematic Modelling}
\label{s:appendix}
In this Appendix, we give details on the kinematic modelling of AzTEC-1 with the $^{\mathrm{3D}}$BAROLO software. $^{\mathrm{3D}}$BAROLO fits 3D tilted-ring models directly to emission-line datacubes, reducing the impact of the beam smearing effect on the derived kinematical parameters \citep[see][for details]{2015MNRAS.451.3021D}. For the kinematic modelling, we use the ALMA datacube of the CO (4-3) emission-line. We fix the geometry of the galaxy to the best-fit parameters found by \citet{2018Natur.560..613T}: we assume a kinematic centre (RA$_\mathrm{c}$, Dec$\mathrm{_c}$) = ($09^\mathrm{h}59^\mathrm{m}42.85^\mathrm{s}$, $+02^{\circ}29'38.23''$), an inclination angle of the galaxy disc with respect to the line of sight $i=44^\circ$ and a position angle of the receding part of the galaxy major axis $\phi = 296^\circ$ (measured counterclockwise from the North direction). A mask is built by smoothing the ALMA CO datacube to a resolution of 0.2$''$ and by running the source finding algorithm on the smoothed datacube with a signal-to-noise cut of 2.5. During the modelling procedure, we use a ring width of 0.04$''$, about half the FWHM of the beam, and we fit the rotation velocity and velocity dispersion only.

\begin{figure*}
\includegraphics[width=1.0\linewidth]{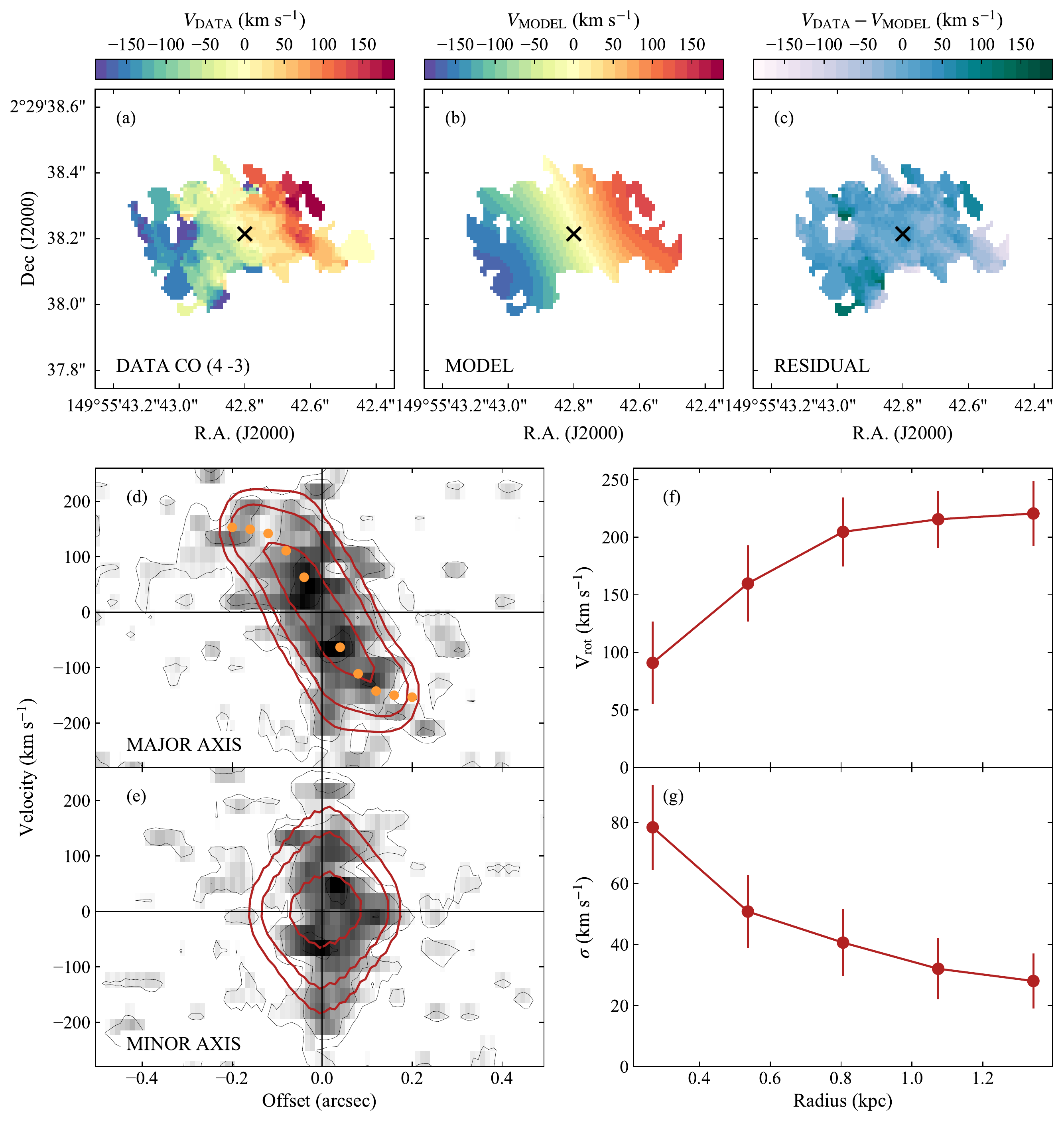}
\caption{Kinematic modelling of AzTEC-1 with $^{\mathrm{3D}}$BAROLO. Panels (a)-(b)-(c): observed velocity field (same as Figure  \ref{fig:figure1}), model velocity field and residuals (data-model). Black crosses represent the kinematic centre of the galaxy. Panels (d)-(e): position-velocity cuts taken along the major and minor axes of the galaxy. Data are shown in greyscale and black contours, model in red contours. Contour levels are at 1.5$\sigma_\mathrm{RMS}$, 3$\sigma_\mathrm{RMS}$ and 5$\sigma_\mathrm{RMS}$, with $\sigma_\mathrm{RMS}=78\,\mu$Jy being the rms noise of the data. Orange dots in panel (d) denote the derived rotation velocity (not corrected for inclination). Panels (f)-(g): rotation curve and velocity dispersion profile along the line of sight.}
\label{fig:appendix}
\end{figure*}

Figure \ref{fig:appendix} shows our best-fit model compared to the observations. Panels (a), (b) and (c) denote the data, model and residual velocity fields, respectively. Panels (d) and (e) show position-velocity diagrams extracted along the major and minor axes, respectively. The data are in greyscale and black contours, the model is represented with red thick contours. Overall, the model traces the data reasonably well. The high-velocity CO emission visible near the centre of the galaxy, which is not reproduced by our simple rotating model, may be due to the presence of strong non-circular motions and/or a starburst-driven outflow. Panel (f) is the rotation curve (inclination-corrected), panel (g) the velocity dispersion profile. We note that our rotation velocity of $\sim 220$ km s$^{-1}$ in the external regions is in good agreement with the maximum rotation velocity of 227 km s$^{-1}$ quoted in \citet{2018Natur.560..613T}. However, we find an average intrinsic gas velocity dispersion of $\sim 50$ km s$^{-1}$, a value slightly lower than the 74 km s$^{-1}$ found by \citet{2018Natur.560..613T}. This discrepancy might be due to the different techniques used to fit the kinematics of the galaxy as well as to the uncertainties related to the large velocity channel width ($\sim30$ km s$^{-1}$) of the ALMA data.

\bsp	
\label{lastpage}
\end{document}